\documentclass[prd,superscriptaddress,floatfix,nofootinbib,superscriptaddress,12pt,titlepage]{revtex4-2} 
\pdfoutput=1 
\usepackage[T1]{fontenc}
\usepackage{amsmath}
\usepackage{amssymb}
\usepackage{amsfonts}
\usepackage{graphicx}
\usepackage{enumitem}
\usepackage{bm}
\usepackage{rotating}
\usepackage{hyperref}
\usepackage{pbox}
\usepackage[dvipsnames]{xcolor}
\usepackage{array}
\usepackage{physics}
\usepackage{dsfont}
\usepackage{mathtools}
\usepackage{leftidx}
\usepackage{lmodern}
\usepackage[normalem]{ulem}
\usepackage{braket}
\usepackage{tensor}
\usepackage{mathrsfs}
\usepackage{float}
\usepackage{dsfont}
\usepackage[margin=1in]{geometry}
\usepackage[title]{appendix}
\usepackage{tcolorbox}

\usepackage{setspace}
\usepackage{tikz}
\usetikzlibrary{arrows.meta,decorations.pathmorphing,math}

\renewcommand{\tilde}{\widetilde}
\newcommand{\skri}{\mathscr{I}}

\newcommand{\sx}{\mathsf{x}}
\newcommand{\sy}{\mathsf{y}}

\newcommand{\bk}{{\bm{k}}}

\newcommand{\supp}{\text{supp}}

\newcommand{\R}{\mathbb{R}}
\newcommand{\C}{\mathbb{C}}
\newcommand{\M}{\mathcal{M}}

\newcommand{\A}{\mathcal{A}}
\newcommand{\W}{\mathcal{W}}
\newcommand{\CS}{C^\infty_c(\M)}
\newcommand{\Sol}{\mathsf{Sol}}

\begin{document}

\title{Holographic reconstruction of asymptotically flat spacetimes}

\author{Erickson Tjoa\footnote{Corresponding author: \href{mailto:e2tjoa@uwaterloo.ca}{e2tjoa@uwaterloo.ca}}}
%\email{e2tjoa@uwaterloo.ca}
\affiliation{Department of Physics and Astronomy, University of Waterloo, Waterloo, Ontario, N2L 3G1, Canada}
\affiliation{Institute for Quantum Computing, University of Waterloo, Waterloo, Ontario, N2L 3G1, Canada}

\author{Finnian Gray \footnote{\href{mailto:fgray@perimeterinstitute.ca}{fgray@perimeterinstitute.ca}}}%
\affiliation{Department of Physics and Astronomy, University of Waterloo, Waterloo, Ontario, N2L 3G1, Canada}
\affiliation{Perimeter Institute for Theoretical Physics, Waterloo, Ontario N2L 2Y5, Canada}

\date{March 30, 2022}

\begin{abstract}
\begin{center}
    -----------------------------------------
\end{center}

\noindent We present a ``holographic'' reconstruction of   bulk spacetime geometry using correlation functions of a massless field living at the ``future boundary'' of the spacetime, namely   future null infinity $\skri^+$. It is holographic in the sense that there exists a one-to-one correspondence between correlation functions of a massless field in four-dimensional spacetime $\M$ and those of another massless field living in three-dimensional null boundary $\skri^+$. The idea is to first reconstruct the bulk metric $g_{\mu\nu}$ by ``inverting'' the bulk correlation functions and re-express the latter in terms of boundary correlators via the correspondence. This effectively allows asymptotic observers close to $\skri^+$ to reconstruct the deep interior of the spacetime using only correlation functions localized near $\skri^+$. 

\vspace{1cm}

\begin{center}
    \emph{“Essay written for the Gravity Research Foundation 2022 Awards for Essays on Gravitation.”}
\end{center}
\end{abstract}

\maketitle
\flushbottom

\section{Introduction}

In most settings, all observers and experimenters are located far away from any astrophysical objects such as black holes or neutron stars. We can therefore regard these observers as \textit{asymptotic observers} infinitely far away from these objects. Such asymptotic observers should be considered to live close to future null infinity\footnote{Note that physical observers do not have access to the usual charges calculated at spatial infinity $i^0$.}  $\skri^+$, because experimentally we \emph{do} detect gravitational and electromagnetic radiation from distant sources. If there is any sense (at all) in which the universe is holographic,   asymptotic observers should be able to learn a lot about the geometry in the {deep interior} of the spacetime without probing every point in the bulk geometry.

One of the less well-known but nonetheless remarkable results in algebraic quantum field theory (AQFT) is that there is a form of \textit{bulk-to-boundary correspondence} between a massless quantum field living on the bulk geometry and another massless quantum field living in its null boundary \cite{Dappiaggi2005rigorous-holo,Dappiaggi2008cosmological,Dappiaggi2009Unruhstate,dappiaggi2015hadamard}. This correspondence is holographic in the sense that the bulk geometry is four-dimensional while the null boundary, which is  future null infinity $\skri^+$ (and possibly any black hole or cosmological horizons), is three-dimensional.  This holography is arguably less lustrous compared to the well-known gauge/gravity duality and AdS/CFT correspondence \cite{Maldacena1999holography,witten1998anti,Gubser1998gaugestring} since it does not involve a strong/weak duality\footnote{That is, one side of the correspondence is strongly interacting (typically the quantum field theory) and the other side weakly interacting (typically the gravitational theory) --- see, e.g., the review ref.~\cite{Hubeny2015review}.}; furthermore, the one-to-one correspondence is between two \textit{non-interacting} quantum fields, not between a gravitational field and a quantum field. {For convenience, we will call this bulk-to-boundary correspondence  \textit{modest holography}, to distinguish it from AdS/CFT or gauge/gravity duality.}

On the other hand, a noteworthy result in the relativistic quantum information (RQI) community shows that local classical spacetime information (metric and curvature) is contained in the correlation functions of the quantum field~\cite{Kempf2016curvature,Kempf2021replace} \emph{and} can be recovered with Unruh-DeWitt (UDW)~\cite{Unruh1979evaporation,DeWitt1979} quantum detectors using physical protocols~\cite{perche2021geometry,perche2022spacetime}.

In this essay we will show that we can reconstruct ``holographically'' the full bulk spacetime geometry using the boundary correlation functions of a quantum field via modest holography {--- in effect combining these orthogonal results.} The scheme is to ``invert'' the bulk correlator to reconstruct the metric, and then with modest holography we effectively reconstruct the metric using the boundary correlators. These boundary correlators can be accessed by asymptotic observers near $\skri^+$, hence the reconstruction can be fully performed by asymptotic observers.

Let us illustrate this using massless scalar fields as an example. Consider a  globally hyperbolic asymptotically flat\footnote{In this essay we assume that the null boundary is $\skri^+$, i.e., asymptotically flat spacetime with future timelike and null infinity \cite{Dappiaggi2005rigorous-holo}.} spacetime $(\M,g_{\mu\nu})$ where $g_{\mu\nu}$ is the metric tensor. The scalar field is assumed to be conformally coupled to gravity. Roughly speaking, in the algebraic framework the scalar field theory is specified by the following {constructions}:

\begin{enumerate}[leftmargin=*,label=(\alph*)]
    \item A unital $*$-algebra $\A(\M)$ generated by \textit{smeared field operator}\footnote{\label{FN:Weyl}Note that in order to avoid ``domain issues'', sometimes it is preferable to work with the Weyl $C^*$-algebra $\W(\M)$, which roughly speaking is generated by the ``exponentiated'' field operators, formally written as $e^{i\hat\phi(f)}$. This can be made more precise (see, e.g., \cite{Khavkhine2015AQFT, fewster2019algebraic}).} $\hat\phi(f)$, where $\hat\phi(f) = \displaystyle\int\dd^4\sx\sqrt{-g}\hat\phi(\sx)f(\sx)$ and $f\in C^\infty_c(\M)$ is a smooth compactly supported function on $\M$. Here we use $\sx$ to label spacetime points.
    
    \item The \textit{canonical commutation relations} (CCR), given by
    \begin{align}
        [\hat\phi(f),\hat\phi(g)] = iE(f,g)\,,\quad E(f,g)\coloneqq \int\dd^4\sx\,\dd^4\sx'\sqrt{-g}\sqrt{-g'}f(\sx)g(\sx')E(\sx,\sx')\,,
    \end{align}
    where $E(\sx,\sy)$ is the \textit{causal propagator} (advanced-minus-retarded Green's function). The causal propagator implements equation of motion, in the sense that we require $(Ef)(\sx) \coloneqq \displaystyle \int \dd^4\sx'\sqrt{-g'}E(\sx,\sx')f(\sx')$ to be a (weak) solution to the conformally coupled Klein-Gordon equation $\nabla_\mu\nabla^\mu \phi - \frac{1}{6} R\phi = 0 $, where $R$ is the Ricci scalar.
    
    \item An \textit{algebraic state} $\omega$, which is a $\C$-linear functional $\omega:\A(\M)\to \C$ such that for any element $A\in \A(\M)$ we have $\omega(A^\dagger A) \geq 0$ and $\omega(\openone) = 1$. 
\end{enumerate}
The $*$-algebra $\A(\M)$ together with CCR forms what is known as an \textit{algebra of observables} for the bulk scalar field\footnote{There are other technical conditions (e.g., the time-slice axiom) but we will not need them in the following discussions.}. In particular, condition (c) suggests that the fundamental objects of the theory are really the \textit{expectation values} of the observables.

The connection to canonical quantization is obtained using the so-called \textit{Gelfand-Naimark-Segal} (GNS) reconstruction theorem \cite{Khavkhine2015AQFT,fewster2019algebraic}. Loosely speaking, it says that for a given $\A(\M)$ and algebraic state $\omega$, we can construct a Hilbert space representation $\pi_\omega:\A(\M)\to\mathcal{H}_\omega$ so that $\ket{0_\omega}\in\mathcal{H}_\omega$ is a vacuum state, uniquely specified by the vacuum two-point functions
\begin{align}
    \omega(\hat\phi(f)\hat\phi(g)) = \braket{0_\omega|\pi_\omega(\hat\phi(f))\pi_\omega(\hat\phi(g))|0_\omega}\,.
\end{align}
The vacuum state is invariant under the full Killing symmetries of $\M$ (in Minkowski space it will be Poincar\'e-invariant). Thus $\hat\phi$ in canonical quantization is a shorthand for the unsmeared field operator $\pi_\omega(\hat\phi(\cdot))$ acting on the Hilbert space $\mathcal{H}_\omega$, 
\begin{align}
    \hat\phi(\sx) \equiv \int\dd^3\bk\,\hat a_\bk u_\bk(\sx) + \hat a_\bk^\dagger u^*_\bk(\sx)\,,
\end{align}
where $\{u_\bk(\sx)\}$ are the positive-frequency mode functions of the Klein-Gordon equation with ladder operators obeying $[\hat a_\bk,\hat a_{\bk'}^\dagger]=\delta^3(\bk-\bk')$. In what follows we will write Hilbert space representation of the field operators $\pi_\omega(\hat\phi(f))$ as simply $\hat\phi(f)$.

In order to study scalar field theory on $\skri^+$, $(\M,g_{\mu\nu})$ is embedded into another conformally related spacetime $(\tilde{\M},\tilde{g}_{\mu\nu})$, in the usual manner~\cite{wald2010general,friedrich1986purely}, with the property that $\tilde{g}_{\mu\nu}=\Omega^2g_{\mu\nu}$ for some nonvanishing smooth function $\Omega>0$ and $\skri^+$ is properly a \textit{null boundary} of $\widetilde{\M}$, so that $(\skri^+,q_{ab})$ will be a null submanifold of $\tilde{M}$ with degenerate metric $q_{ab}=\tilde{g}_{\mu\nu}|_{\skri^+}$. The boundary scalar theory differs considerably from the bulk scalar theory in that $\skri^+$ is a null hypersurface, and that there is no equation of motion (hence no causal propagator) that induces a CCR algebra at $\skri^+$. The canonical quantization is thus far from obvious, but the algebraic framework naturally accounts for this possibility. The basic idea is that since massless fields in the bulk can reach $\skri^+$ (if we think of it in terms of the unphysical spacetime $\tilde{M}$), the algebraic quantization only needs to ensure that the boundary theory respects the ``projection'' of the bulk solution of the Klein-Gordon equation to $\skri^+$. In practical terms, it means that~\cite{Dappiaggi2005rigorous-holo}
\begin{enumerate}[label=(\alph*),leftmargin=*]
    \item The algebra of observables $\A(\skri^+)$ is generated by \textit{smeared boundary field operators} $\hat\varphi(\psi)$, where $\psi\in \Sol_\R(\skri^+) = \{\psi\in \C^{\infty}_c(\M): \psi,\partial_u\psi\in L^2(\skri^+,\dd u\,\dd^2x^A)\}$. That is, $\psi$ and $\partial_u\psi$ (its first derivative with respect to the null coordinate $u$) are smooth, compactly supported and square-integrable on $\skri^+$.\footnote{The space $\Sol_\R(\skri^+)$ is actually is upgraded to a symplectic vector space by introducing a symplectic form $ \sigma_{\skri}(\psi_1,\psi_2)=\int_\skri
    \dd u\,\dd^2 x^A(\psi_2\partial_u\psi_1 -\psi_1\partial_u\psi_2)$ from which the Weyl algebra follows~\cite{bratteli2002operatorv2} (see footnote~\ref{FN:Weyl})} The natural coordinate system on $\skri^+$ is provided by a \textit{Bondi chart} $(u,x^A)$, where $x^A$ labels points on the 2-sphere $S^2$.
    
    \item If $Ef\in\Sol_\R(\M)$ and $Eg\in \Sol_\R(\M)$ are real solutions to the Klein-Gordon equation with compact Cauchy data, then the boundary fields commute whenever $f,g$ are spacelike-separated in $\M$. That is, $[\hat\varphi(\psi_f),\hat\varphi(\psi_g)] = 0$, where $\Gamma:\Sol_\R(\M)\to \Sol_\R(\skri^+)$ is a projection map to $\skri^+$ and
        \begin{align}
            \psi_f =\Gamma (Ef)= \lim_{\skri^+}  \,\Omega^{-1}Ef\,.
        \label{eq: boundary-data}
        \end{align}
    
    \item The algebraic state $\omega_\skri:\A(\skri^+)\to \C$ is defined the same way as before, but the GNS representation induces a vacuum state $\ket{0_\skri}$ that is invariant under the action of the \textit{asymptotic} symmetry group of asymptotically flat spacetimes, namely the \textit{Bondi-Metzner-Sachs} group (denoted $\mathsf{BMS}$).
\end{enumerate}

Since the boundary fields are projections of the bulk fields, there is a sense in which the algebra of observables $\A(\skri^+)$ is much larger than $\A(\M)$. In technical terms, we say that there exists an injective and isometric $*$-homomorphism between the two $*$-algebras $i:\A(\M)\to\A(\skri^+)$. Most importantly, the algebraic state $\omega_\skri$, which induces the $\mathsf{BMS}$-invariant vacuum state $\ket{0_\skri}$ in the GNS representation, can be ``pulled back'' using the $*$-homomorphism to \textit{uniquely} define a bulk state $\omega\coloneqq i^*\omega_\skri:\A(\M)\to\C$, with the property that the expectation values in $\M$ and in $\skri^+$ are compatible. That is, if $A\in \A(\M)$ then\footnote{In practice, $A$ could be a product of several smeared field operators, so this covers arbitrary $N$-point functions of the theory.}
\begin{align}
    (i^*\omega_\skri)(A) = \omega_\skri(i(A))\,.
    \label{eq: algebraic-VEV-holography}
\end{align}
The $*$-homomorphism action on generators of $\A(\M)$ is given by $i(\hat\phi(f)) = \hat\varphi(\Gamma Ef)$ for every $ f\in \CS $. If $\M$ is Minkowski space, the pullback $i^*\omega_\skri$ \textit{uniquely}~\cite{Moretti2005BMS-invar,Moretti:2006SP-invar} defines the familiar \textit{Minkowski vacuum} $\ket{0_\textsc{M}}$: the $\mathsf{BMS}$-invariant vacuum two-point functions in $\skri^+$ induces a unique Poincar\'e-invariant vacuum two-point functions. {Speaking generally, the vacuum state defined by this pullback will be invariant under any symmetries of the bulk spacetime \cite{Moretti:2006SP-invar}.}

We are now ready to perform holographic reconstruction of the bulk geometry. For this we need to work in the GNS representation of the bulk and boundary fields. The corresponding vacuum two-point function at the bulk and the boundary can be written as 
\begin{align}
    \mathsf{W}_\M(f,g)\coloneqq \braket{0_\M|\hat\phi(f)\hat\phi(g)|0_\M}\,,\quad \mathsf{W}_\skri(\psi_f,\psi_g)\coloneqq \braket{0_\skri|\hat\varphi(\psi_f)\hat\varphi(\psi_g)|0_\skri}\,,
\end{align}
where $\psi_f = \Gamma Ef$. The modest holography \eqref{eq: algebraic-VEV-holography} says that the correlators agree~\cite{Dappiaggi2005rigorous-holo}:
\begin{align}
    \mathsf{W}_\M(f,g) = \mathsf{W}_\skri(\psi_f,\psi_g)
    \,.
\end{align} 
Since $\skri^+$ is a \textit{universal structure} of all asymptotically flat spacetimes,
modest holography dictates that the calculation for two distinct asymptotically flat spacetimes $\M_1,\M_2$ differs in the projection map $\Gamma$. This is so because the corresponding unphysical spacetimes $\tilde\M_1,\tilde\M_2$ embed $\skri^+$ as their boundaries differently (i.e., the conformal factors are different).

The next step is to show that we can reconstruct the metric using the \textit{bulk} correlator $\mathsf{W}_\M(f,g)$. This already follows from the work of Saravani, Aslanbeigi and Kempf \cite{Kempf2016curvature,Kempf2021replace}, who showed that in effect we can recover the metric in (3+1) dimensions from the correlation functions:
\begin{align}
    g_{\mu\nu}(\sx) = {-\frac{1}{8\pi^2}}\lim_{\sy \to \sx}\partial_\mu\partial_\nu \mathsf{W}_\M(\sx,\sy)^{-1}\,.
    \label{eq: metric-from-wightman}
\end{align}
The reason why this works is that all physically reasonable algebraic states of the field theory are required to be \textit{Hadamard states}. That is, if $\sx,\sy\in \M$ are two spacetime events that are sufficiently close, then the \textit{unsmeared} two-point functions take the form \cite{wald1994quantum}
\begin{align}
    \mathsf{W}_\M(\sx,\sy) = \frac{U(\sx,\sy)}{{8\pi^2}\sigma_\epsilon(\sx,\sy)} + V(\sx,\sy)\log(\sigma_\epsilon(\sx,\sy)) + Z(\sx,\sy)\,,
\end{align}
where $U,V,Z$ are regular smooth functions. The bi-scalar $\sigma_\epsilon(\sx,\sy)$ is the Synge world function augmented with $i\epsilon$ prescription (because of its distributional nature of $\mathsf{W}_\M(\sx,\sy)$): {that is,}
\begin{align}
    \sigma_\epsilon(\sx,\sy) &= \sigma(\sx,\sy) +2i\epsilon(T(\sx)-T(\sy)) +\epsilon^2\,,\\
    \sigma(\sx,\sy) &= \frac{1}{2}(\tau_\sy-\tau_\sx)\int_\gamma g_{\mu\nu}(\lambda)\dot{\gamma}^\mu(\lambda)\dot{\gamma}^\nu(\lambda)\dd\lambda\,,
\end{align}
where  $\sigma(\sx,\sy)\equiv \sigma_{\epsilon=0}(\sx,\sy)$ is the Synge world function, $T$ is a global time function (this always exists because we assume $\M$ to be globally hyperbolic) and $\gamma(\tau)$ is a geodesic curve with affine parameter $\tau$ with $\gamma(\tau_\sx)=\sx$ and $\gamma(\tau_\sy)=\sy$. Schematically, Eq.~\eqref{eq: metric-from-wightman} follows from the fact that when $\sy\approx \sx$ we have $\Delta\sx=\sx-\sy\approx 0$ and 
\begin{align}
    \mathsf{W}_\M(\sx,\sy)^{-1}\approx {8\pi^2}\sigma(\sx,\sy) \sim {4\pi^2} g_{\mu\nu}(\sx) \Delta x^\mu \Delta x^\nu + O(\Delta\sx^2)\,.
\end{align}
\indent In principle, since $\mathsf{W}_\M(\sx,\sy)$ is distributional in nature, we should work with smeared two-point functions $\W_\M(f,g)$ with $\sx\in \supp(f)$ and $\sy\in \supp(g)$. In other words, there is a ``resolution limit''\footnote{For closely separated $\sx,\sy$ the reciprocal $\mathsf{W}_\M^{-1}$ is approximately proportional to $\sigma(\sx,\sy)$ which is a regular function. Thus we can set $\epsilon=0$ and compute \eqref{eq: metric-from-wightman} directly without smearing functions to get the metric.}  defined by the supports of the smearing functions $f,g$: physically, it means that vacuum noise prevents us from reconstructing the metric with infinite accuracy. Taking this into account, we calculate the metric using finite differencing: pick $f,g$ to be sharply peaked functions with characteristic widths $a_0$ localized around $\sx$ and $\sy$ respectively\footnote{We can take $f,g$ to be Gaussian as an approximation since the tails quickly become negligible and are effectively compactly supported. Then $a_0$ measures the width of the Gaussian (say, several standard deviations).}. The finite-difference approximation of $\partial_\mu\partial_{\nu'}\mathsf{W}(\sx,\sx')^{-1}$ applied to the reciprocal of the Wightman function reads
\begin{align}
    \partial_\mu\partial_{\nu'}\mathsf{W}(\sx,\sx')^{-1}
    &\approx \frac{\mathsf{W}(\sx+\epsilon^\mu,\sx'+\epsilon^\nu)^{-1}-\mathsf{W}(\sx+\epsilon^\mu,\sx')^{-1}}{\delta^2}-\frac{\mathsf{W}(\sx,\sx'+\epsilon^\nu)^{-1}-\mathsf{W}(\sx,\sx')^{-1}}{\delta^2}\,.
\end{align}
Here $\epsilon^\mu$ points in the direction of coordinate basis $\partial_\mu$ with norm $\sqrt{|\epsilon^\mu\epsilon_\mu|} = \delta \ll 1$. A change of variable (shift by $\epsilon^\mu$) and smearing the Wightman functions before taking its reciprocal allows us to write the metric approximation as
\begin{align}
    g_{\mu\nu}(\sx) &\approx {-\frac{1}{8\pi^2\delta^2}}\Bigr[\mathsf{W}(f_\epsilon,g_\epsilon)^{-1} - \mathsf{W}(f_\epsilon,g)^{-1} -\mathsf{W}(f,g_\epsilon)^{-1}+\mathsf{W}(f,g)^{-1}\Bigr]\,,
    \label{eq: metric-reconstruction}
\end{align}
where $f_\epsilon(\sx)=f(\sx-\epsilon^\mu)$ and $g_\epsilon(\sx) = g(\sx-\epsilon^\nu)$. The approximation gets better with smaller $\delta$ but this is bounded below by the resolution provided by characteristic widths of $f,g$. Note that the spacetime smearing functions must be properly normalized to reproduce the metric.

The final step to obtain the holographic reconstruction is to combine this step from Saravani, Aslanbeigi and Kempf \cite{Kempf2016curvature,Kempf2021replace} with the modest holography---{crucially, the state induced in the bulk by \eqref{eq: algebraic-VEV-holography} is automatically Hadamard~\cite{Moretti:2006SP-invar}}. That is, using Eq.~\eqref{eq: algebraic-VEV-holography} together with \eqref{eq: metric-reconstruction} we get
\begin{align}
    g_{\mu\nu}(\sx) &\approx -{\frac{1}{8\pi^2\delta^2}}\Bigr[\mathsf{W}_\skri(\psi_{f_\epsilon},\psi_{g_\epsilon})^{-1} - \mathsf{W}_\skri(\psi_{f_\epsilon},\psi_{g})^{-1}-\mathsf{W}_\skri(\psi_{f},\psi_{g_\epsilon})^{-1}+\mathsf{W}_\skri(\psi_{f},\psi_{g})^{-1}\Bigr]\,.
    \label{eq: metric-reconstruction-boundary}
\end{align}
As before $\psi_{f_\epsilon}=\Gamma (Ef_{\epsilon})$. Since the RHS is completely given in terms of boundary quantities, this gives us a genuine holographic reconstruction of the bulk interior from the null boundary.

Moreover, we note that while the RHS of \eqref{eq: metric-reconstruction-boundary} is defined for fields living on $\skri^+$, it can be shown that these correlators $\mathsf{W}_\skri(\psi_f,\psi_g)$ can be constructed by \textit{asymptotic observers} \textit{near} $\skri^+$ \cite{tjoa2022holography}. The idea is that while no physical observers can follow null geodesics exactly on $\skri^+$, we can perform a large-$r$ expansion of the bulk field operator (see, e.g., \cite{strominger2018lectures}). The asymptotic observers near $\skri^+$ will thus find that the bulk correlation functions $\mathsf{W}_\M(f,g)$ very close to $\skri^+$ are at leading order given exactly by $\mathsf{W}_\skri(\psi_f,\psi_g)$ and subleading corrections suppressed by higher powers of $1/r$. These correlators can be measured, for instance, using the UDW detector formalism \cite{Unruh1979evaporation,DeWitt1979,Pipo2018direct}.

In the case where the bulk geometry contains a black hole, such as the Schwarzschild geometry, the null boundary needs to be extended to include the future horizon $\mathscr{H}^+$, since on the exterior geometry $\skri^+$ alone does not provide enough Cauchy data \cite{Dappiaggi2005rigorous-holo,Dappiaggi2008cosmological,Dappiaggi2009Unruhstate,dappiaggi2015hadamard}. It would be interesting to see how much asymptotic observers (and possibly together with near-horizon observers) can learn about the black hole in the bulk using the correlators associated with the Unruh vacuum and whether it is possible to reconstruct at least the exterior geometry from the null boundary $\mathscr{H}^+\cup\mathscr{I}^+$. Furthermore, this modest holography would also be of practical computational interest in semiclassical calculations: it may give us a way to greatly simplify extremely complicated calculations of correlation functions in curved spacetimes (see, e.g., \cite{Casals2020commBH}), since the near-horizon and near-$\skri^+$ correlators are much better behaved.

{Finally, although we have made use only of the properties of ordinary QFT in curved spacetime, these ideas should in principle carry over to the asymptotic quantization of gravity~\cite{Ashtekar1981asymptotic,Ashtekar1981radiative,Ashtekar2018infraredissues}, and provide a new direction to explore the key differences arising from the nature of the gravitational field.}

\noindent \textbf{Acknowledgments.} We thank Robert Mann and David Kubiz\v{n}\'ak for valuable comments on the manuscript. E.T. acknowledges generous support of Mike and Ophelia Lazaridis Fellowship.  F.G. is funded from the Natural Sciences and Engineering Research Council of Canada (NSERC) via a Vanier Canada Graduate Scholarship. This work was also partially supported by NSERC and partially by the Perimeter Institute for Theoretical Physics. Research at Perimeter Institute is supported in part by the Government of Canada through the Department of Innovation, Science and Economic Development Canada and by the Province of Ontario through the Ministry of Colleges and Universities. Perimeter Institute, Institute for Quantum Computing and the University of Waterloo are situated on the Haldimand Tract, land that was promised to the Haudenosaunee of the Six Nations of the Grand River, and is within the territory of the Neutral, Anishnawbe, and Haudenosaunee peoples.

\bibliography{peeling}

\end{document}